\documentstyle[euromacr,amssymb,psfig]{europhys}

\begin{document}
\shorttitle{Helix-Coil transition in grafted chains}
\title{On the Helix-Coil transition in grafted chains}
\author{A. Buhot\footnote{Present address : Theoretical Physics,
1 Keble Road, Oxford OX1 3NP, U.K.} and A. Halperin}
\institute{UMR 5819, DRFMC/SI3M, CEA-Grenoble, 17 , rue des Martyrs,\\
38054 Grenoble Cedex 9, France}
\date{\today}
\pacs{
\Pacs{61}{25.Hq}{Macromolecular and polymer solutions; polymer melts;
swelling}
\Pacs{87}{15.-v}{Biomolecules: structure and physical properties}
\Pacs{87}{15.He}{Dynamics and conformational changes}
}
\maketitle

\begin{abstract}
The helix-coil transition is modified by grafting to a surface. This
modification is studied for short peptides capable of forming 
$\alpha$-helices. Three factors are involved: (i) the grafting can 
induced change of
the boundary free energy of the helical domain (ii) the van der Waals
attraction between the helices and (iii) the crowding induced stretching of
the coils. As a result the helix-coil transition acquires ``all or nothing''
characteristics. In addition the transition temperature is elevated and the
transition itself sharpens as the grafting density increases.
\end{abstract}

During the past two decades the physics of polymer brushes formed by
terminally anchored chains were studied 
extensively\cite{Milner,HTL,IGAL}. Most of the research effort 
dealt with brushes of flexible,
synthetic polymers, devoid of internal degrees of freedom. In contrast, this
letter concerns brushes formed by biopolymers capable of undergoing a
cooperative helix-coil transition. It is motivated by two experimental
observations. First, the promotion of the adhesion and spreading of cells by
brushes of collagen model peptides\cite{Tirrell}. Second, membrane fusion
induced by model fusogenic peptides grafted to vesicles\cite{Fusion}. In
both cases, the function of the short peptide chains was correlated with a
helical state. Furthermore, the helical state was favored by the grafting.
With this in mind we present a highly simplified theoretical model for the
helix-coil transition in brushes of short, laterally immobile peptides. In
particular, we discuss the transition temperature $T_{t}$ and the width of
the transition $\Delta T$ as a function of grafting density, $\Sigma^{-1}$.
We focus on the simplest situation, of short neutral homopeptides forming a
single stranded $\alpha$-helix. As we shall see, the grafting of the chains
can lead to qualitative modifications of the helix-coil transition: (i) in
marked distinction to the case of a free peptide, the helix-coil transition
in an isolated grafted peptide can acquire ``all or nothing''
characteristics; (ii) $T_{t}$ is elevated with the grafting density while 
$\Delta T$ decreases; (iii) eventually, for high grafting density the
transition may take place as a first-order phase transition. These
distinctive features arise because of three factors: The lower
configurational entropy of the monomer at the grafting site favors helix
formation even in isolated chains. In a brush, the helical state is also
promoted by the crowding induced stretching and by the attractive van der
Waals interaction between the helices.

In a $\alpha$-helix \cite{BP,PS,GK} the $i$th monomer forms $H$-bonds 
with the $(i-3)$th and the $(i+3)$th monomers. Overall, a helical
domain consisting of $n$ monomers contains $n-2$ $H$-bonds. It is convenient
to consider the chain as a sequence of bonds. Choosing the coil state as a
reference state, each helical bond is associated with an excess free energy 
$\Delta f$ reflecting the formation of $H$-bonds, the accompanying change in
solvation and the loss of configurational entropy. $\Delta f$ is a function
of the temperature, $T$. $\Delta f$ vanishes at the transition temperature 
$T_{\ast}$, while $\Delta f<0$ when $T<$ $T_{\ast}$ and $\Delta f>0$ for $T>
T_{\ast}$. The terminal bonds, at the boundary of the helical domains,
are associated with an additional free energy penalty $\Delta f_{t}$. This
arises because the terminal bonds lose their configurational entropy but do
not contribute $H$-bonds. $\Delta f_{t}$ plays the role of an interfacial
free energy associated with the helix-coil boundary. It is customary to
formulate the theory of the helix-coil transition in terms of the Bragg-Zimm
parameters $s=\exp (-\Delta f/kT)$ and $\sigma =\exp (-2\Delta f_{t}/kT)$. 
$\sigma$ is independent of $T$ and is typically of order of 
$10^{-3}-10^{-4}$,
depending on the identity of the amino acid residues forming the peptide.
On the other hand, $s$ is a function of $T$ and in the vicinity of 
$T_{\ast}$ it varies as $s-1\sim (T_{\ast}-T)/T_{\ast}$. 
In an infinite chain, the
plot of the fraction of helical bonds, $\theta$ vs. $s$ is sigmoid and the
width of the transition is $\Delta T\sim T_{\ast}\sigma^{1/2}$. Since the
chain is one dimensional object, a first order phase transition is
impossible. As a result the chain consists of an alternating sequence of
helical and coil regions. The minimal size of a domain comprising a helical
and a coil regions, as obtained at $T_{\ast}$, is $\sigma^{-1/2}$. When the
polymerization degree, $N$ is much larger than $\sigma^{-1/2}$ the chain may
be considered as infinite. It consists of a large number of domains and the
width of the transition, $\sigma^{1/2}$, is independent of $N$. On the
other hand, when $N\leq \sigma^{-1/2}$ the chain incorporates typically only
one helical region and the width of the transition is of order $1/N$. The
statistical physics of such short oligopeptides are well described by the
``one sequence approximation'' where the chain is assumed to contain a
single helical region \cite{PS}. The customary formulation of this
approximation utilizes the appropriate partition function. For the purposes
of our discussion it is convenient to utilize the corresponding free energy.
The chemical potential of a chain supporting a helical domain consisting of 
$n$ bonds is $\mu_{0}(n)=n\Delta f+2\Delta f_{t}-kT\ln (N-n-1)$ where the
last term is the entropy associated with the placement of the helical
segment along the chain. Altogether there are $N-2$ bonds and the first
monomer of the helical segment of length $n$ can be placed in any of the 
$(N-2)-n+1$ sites. For $n=0$ we have $\mu_{0}(0)=0$. The fraction of chains
with a given $n$, $p_{n}$, is determined by minimizing $\Omega =
\Omega_{0}+M\sum\nolimits_{n}p_{n}[\mu_{0}(n)+kT\ln p_{n}]$ subject to the
constraint $\sum\nolimits_{n}p_{n}=1$. This leads to $\mu =\mu_{0}(n)+kT\ln
p_{n}=const^{\prime}$ since all species coexist in equilibrium. For our
discussion it is sufficient to consider the dominant term in $\Omega$, as
specified by $\partial \mu_{0}(n)/\partial n=0$. When $\Delta f\geq 0$ the
minimal $\mu_{0}(n)$ is $\mu_{0}(0)=0$ and the majority of chains contain
no helical region. On the other hand, for $\Delta f<0$ the probability
distribution peaks at $n_{\ast}=N-1-\mid kT/\Delta f\mid$. Accordingly, 
$n_{\ast}$ attains its maximum value, $n_{\ast}=N-2$, when $\mid \Delta
f\mid =kT$. The discussion as presented above applies to free chains. Two
features are of special importance for future reference: (i) $\Delta f_{t}$
is independent of the position of the helical region and, as a result (ii)
all placements of the helical segments are equally probable. Because of the
associated entropy $n_{\ast}$ is lower than $N-2$ for $0>\Delta f>-kT$.

The situation described above is modified significantly when the chain is
grafted, terminally anchored, to a surface. We first discuss the case of a
single chain grafted to a flat surface. The grafting gives rise to two
effects. First, the overall number of configurations of the chain is reduced
because of the presence of an impenetrable wall. As a result, the overall
configurational entropy is reduced by $\Delta S\sim \ln N$\cite{EE}. This is
expected to lead to a small changes in $s$ and $\sigma $ which we will
ignore since they do not give rise to qualitative effects. Second, the two
boundaries of the helical region are no longer equivalent because of the
presence of the surface. The surface effect is expected to decay with the
length of the coil region separating the helical region from the wall. The
maximal effect is attained when the helical region is initiated at the wall.
The penalty of the ``free'' boundary remains $\Delta f_{t}.$ However, the
terminal penalty at the wall, $\Delta f_{w},$ can be different. The
difference can arise from two sources. First a modification of the torsional
potential for the terminal monomer and the consequent reduction of its
configurational entropy. Second, change in the electrostatic interactions
experienced by the terminal monomer at the grafting site. As a result, the
Bragg-Zimm parameters are modified and for a helical sequence at the wall 
$\sigma_{g}=\exp [-(\Delta f_{w}+\Delta f_{t})/kT].$ In the following we
consider the physically plausible case of $\Delta f_{w}<\Delta f_{t}$. For
simplicity we assume that $\Delta f_{w}$ is attained only when the helical
sequence begins at the grafting site. In this situation a helical sequence
of length $n$ may assume two states: (i) When the terminal monomer is not at
the wall, the helical sequence can be freely placed at the available
remaining sites. The ``free'' sequence is associated with $\mu_{f0}=n\Delta
f+2\Delta f_{t}-kT\ln (N-n-2)$. (ii) A ``bound'' state when the terminal
monomer is at the grafting site and $\mu_{b0}=n\Delta f+\Delta f_{t}+\Delta
f_{w}$. The two states are equally stable when $\mu_{f0}=\mu_{b0}$. This
condition is satisfied for $n_{\dagger}$ given by $n_{\dagger}=N-2-\exp
[(\Delta f_{t}-\Delta f_{w})/kT]$. For $n>n_{\dagger}$ the bound
state is favored. Accordingly, the bound state is always dominant when 
$n_{\dagger}=1$ or $\ln (N-3)\leq (\Delta f_{t}-\Delta f_{w})/kT$.
Furthermore, among the bound states, the fully helical state is always the
most stable {\em i.e.}, $\mu_{b0}(N)<\mu_{b0}(n)$ whenever $\Delta f<0$.
For this choice of $N$ the coil state remains the most probable state, 
$n_{\ast}=0$ when $\Delta f>0$ while for any $\Delta f<0$ the most probable
state is fully helical, $n_{\ast}=N-2$. Thus, the grafting of short chains
modifies the nature of the helix-coil transition that acquires ``all or
nothing'' characteristics. In the remainder of this letter we focus on
brushes of such chains.

Our earlier considerations concerned a single grafted chain. With them in
mind we model a grafted layer consisting of many chains as a mixture of
chains in a coil state and chains in a fully helical state. In this case,
additional contributions come into play. One is the van der Waals attraction
between the rods. Another important contribution is the crowding induced
extension of the coils in the brush regime. To quantify this picture it is
necessary to introduce further assumptions concerning the orientation
imposed on the helices by the grafting. For simplicity we will consider the
case of grafting sites enforcing perpendicular orientation of the helices
with respect to the surface. We further limit the discussion to the case of
rigid junctions that is, the helices cannot bend at the grafting sites. Thus
far, our considerations were directed at chains with $N\leq 
\sigma^{-1/2}\approx 10^{2}$. At this point it is necessary to 
limit our discussion
further to chains with $N\leq N_{\ast}\approx (a/a_{h})^{2}$ where $a$ is
the effective monomer size in the coil state while $a_{h}=1.5$\AA\ is the
projection of a residue on the axis of the helix. While $a$ depends on the
identity of the residue, $a/a_{h}>1$ and typically $N_{\ast}\gtrsim 50$.
This choice is necessary because the length of a fully helical chain, $L$,
is smaller than the radius of the corresponding random coil, $R_{0}\approx
N^{1/2}a$ when $N\leq N_{\ast}$ while $L>R_{0}$ if $N>N_{\ast}$. Thus, in
one case the helices are submerged in the brush formed by the coils while in
the other the helices can be partially exposed (Figure 1). This choice of
the $N$ range is dictated by the experimental systems motivating this work 
\cite{Tirrell,Fusion}. With the system fully specified we are in a
position to write down the free energy per chain, $F$. For simplicity we use
a modification of the Alexander model \cite{SA}, assuming that the coils are
uniformly stretched and that concentration profile of the monomers of the
coils is step-like. We consider a planar grafted layer such that the area
per chain is $\Sigma$ and the fraction of chains in a helical state is 
$\theta$. $F$ consists of four terms, $F=F_{0}+\theta F_{vdW}+(1-\theta)
F_{cc}+\theta F_{hc}$. $F_{0}$ reflects the contribution of the non
interacting chains $F_{0}\approx \theta (N\Delta f+\Delta f_{t}+\Delta
f_{w})+kT[\theta \ln \theta +(1-\theta )\ln (1-\theta )]$ where the last
term is the mixing entropy associated with the possible placements of chains
in the two possible states. The van der Waals attraction between two
oriented helices is modeled as the interaction between two parallel rods of
diameter $d$. When the distance between two neighboring rods, $D$, is small
in comparison to $L$, their length, $F_{vdW}\sim -L/D^{5}$\cite{JNI} while
in the opposite limit $F_{vdW}\sim -L^{2}/D^{6}$. In the first case, an
element of the rod experiences, in effect, an interaction with an infinite
neighboring rod. In the second limit, $F_{vdW}$ reflects the sum of all
pairwise interactions between the elements of the two rods. The van der
Waals energy of a rod within the grafted layer, where the minimal distance
between two neighboring helices is $(4\Sigma /\pi \theta )^{1/2},$ is 
\begin{figure}
\centerline{\psfig{file=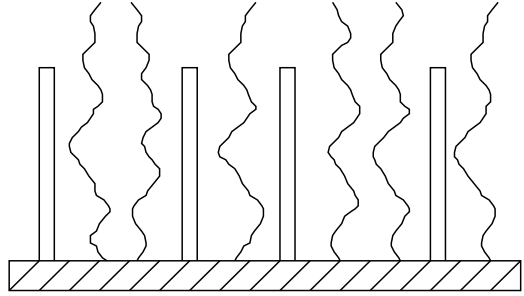,width=8cm}}
\caption{A brush of short chains undergoing a helix-coil transition.
The helical chains are short compared to the span of the coils and are thus
fully embedded in the brush formed by the coils.}
\end{figure}
\begin{equation}
F_{vdW}=-\frac{ALd^{4}}{18}\left( \frac{\pi \theta }{4\Sigma }\right)
^{5/2}\arctan \sqrt{\frac{\pi \theta L^{2}}{4\Sigma }}  \label{1}
\end{equation}
where $A$ is the Hamaker constant. In a ``normal'' brush, the overlap
between the coils gives rise to chain extension along the normal to the
surface. The stretching reduces the number of monomer-monomer contacts at
the price of an extension penalty. In our situation the picture is somewhat
different. First, it is necessary to allow for both coil-coil and coil-helix
interactions. Second, only the coils are extendible. Finally, for our choice
of $N$, the helices are fully immersed in a brush of coils. $F_{cc}$ allows
for the elastic free energy of the coils as well as for monomer-monomer
interactions involving the coils. This contribution has the form of the free
energy of a brush formed by coils that is, $F_{cc}/kT\approx
H^{2}/Na^{2}+v_{cc}N^{2}a^{3}(1-\theta )/H\Sigma$ where $H$ is the
thickness of the brush and $v_{cc}$ is the virial coefficient for
monomer-monomer interactions involving coils\cite{Foot0}. The first term
reflects the Gaussian stretching penalty while the second allows for the
repulsive monomer-monomer contacts. $F_{cc}$ must be supplemented by $F_{hc}$
that reflects the interactions between the coils and the helices, 
$F_{hc}/kT\approx v_{hc}N^{2}a^{3}(1-\theta)/H\Sigma$. Here $v_{hc}$ is the
virial coefficient for the binary interactions between helical monomers and
coil monomers. For simplicity we consider the case of $v_{hc}=v_{cc}=v$. In
this case the equilibrium thickness of the brush $H_{eq}/a\approx
N(va^{2}/\Sigma)^{1/3}$, as specified by $\partial F/\partial H=0$, is
independent of $\theta$ so long as $\theta $ is small enough to ensure
overlap between the coils\cite{Foot1}. For such a choice of $\theta$ the
equilibrium form of the free energy per chain, as obtained by substituting 
$H_{eq}$ into $F$, is 
\begin{eqnarray}
\nonumber \frac{F}{NkT} & \approx & \frac{\theta \Delta f}{kT} + 
\frac{\theta (\Delta f_{t}+\Delta f_{w})}{NkT} - \lambda \theta^{7/2} 
\left( \frac{a^{2}}{\Sigma} \right)^{5/2}
+ (1-\theta)\left(\frac{va^{2}}{\Sigma}\right)^{2/3}\\
& & + \frac{1}{N}\left[ \theta \ln \theta +(1-\theta)\ln (1-\theta)\right] 
\label{2}
\end{eqnarray}
where $\lambda \approx \frac{ALd^{4}}{NkTa^{5}}$. $F_{vdW}$ in (\ref{2}) is
approximated by the $\pi \theta L^{2}/\Sigma \gg 1$ limit of (\ref{1}). As a
result, $F_{vdW}$ is overestimated for $\theta \ll 1$. This does not affect
our analysis since the contribution of the van der Waals attraction in this
regime is negligible. The corresponding spinodal condition, 
$\partial^{2} F/\partial \theta^{2}=0$, leads to $f(\theta) = \theta^{5/2}
(1-\theta) \approx (4/35) (\Sigma/a^{2})^{5/2}(\lambda N)^{-1}$ revealing
a critical point at $\theta_{c}=5/7$ where $f(\theta)$ exhibits a maximum.
The critical grafting density is thus specified by 
\begin{equation}
\Sigma_{c}/a^{2}\approx \lambda^{2/5}N^{2/5}  \label{3}
\end{equation}
and for $\Sigma <\Sigma_{c}$ the helix-coil transition within the layer
proceeds as a first-order phase transition. For lower grafting densities, 
$\Sigma >\Sigma _{c},$ the helix-coil transition is cooperative but no phase
transition is involved. In this last regime it is of interest to
characterize the dependence of the transition temperature $T_{t}$ and the
width of the transition $\Delta T$ on $\Sigma $ and $N$. To this end it is
helpful to recast the equilibrium condition, $\partial F/\partial 
\theta =0$, in the form 
\begin{equation}
\frac{\theta}{1-\theta }=K(\theta)\approx s^{N}\sigma_{g}\exp \left[ 
\frac{7}{2}\lambda N\left(\frac{\theta a^{2}}{\Sigma }\right)^{5/2}
+N \left(\frac{va^{2}}{\Sigma}\right)^{2/3}\right]   \label{4}
\end{equation}
where $K(\theta)$ is the equilibrium constant governing the ratio of
helices and coils. A rough idea concerning the transition is obtained by
identifying it with the condition $\theta =1/2$ i.e., $K(\theta )=1$. It is
helpful to consider first the case of non interacting mushrooms, $\Sigma
\rightarrow \infty$, when $K(\theta )=s^{N}\sigma _{g}$. Since $s$ in the
vicinity of $T_{\ast }$ is given by $s\approx \exp \left[ -r\left(\frac{T}{
T_{\ast}}-1\right) \right]$ where $r$ is a phenomenological constant, this
condition leads to 
\begin{equation}
T_{t}(\infty )\approx T_{\ast}+\frac{T_{\ast}}{Nr}\ln \sigma_{g}\lesssim
T_{\ast}.  \label{5}
\end{equation}
Similarly we define the width of the transition as $\Delta T=T_{-}-T_{+}$
where $T_{+}$ is the temperature for which $\theta_{+}=9/10$ while $T_{-}$
is the temperature corresponding to $\theta_{-}=1/10$\cite{Foot2}. The
ratio $K(\theta_{+})/K(\theta_{-})=\left( s_{+}/s_{-}\right)^{N}=81$ may
be rewritten as $\ln 81\approx Nr\Delta T(\infty)/T_{\ast}$ or 
\begin{equation}
\Delta T(\infty )\approx \frac{T_{\ast}}{Nr}.  \label{6}
\end{equation}
By using the same argument for a brush with $\Sigma \ll R_{0}^{2}$ we obtain 
\begin{equation}
T_{t}=T_{t}(\infty )+\frac{T_{\ast}}{r}\left[ \frac{7}{2}\lambda \left( 
\frac{a^{2}}{2\Sigma }\right)^{5/2}+\left(\frac{va^{2}}{\Sigma}\right)
^{2/3}\right]   \label{7}
\end{equation}
and 
\begin{equation}
\Delta T=\Delta T(\infty )-\frac{7}{2}\frac{T_{\ast}}{r}\lambda \left( 
\frac{9a^{2}}{10\Sigma }\right)^{5/2}.  \label{8}
\end{equation}
As $\Sigma$ decreases, $T_{t}$ increases while $\Delta T$ decreases. In
other words, the stability of the helical state grows and the transition 
becomes sharper when the grafting density increases. 
Eventually, at the critical
point, $\Delta T=0$ thus signaling the onset of a phase transition. Note
that within our model the decrease of $\Delta T$ is due only to $F_{vdW}$
while the increase of $T_{t}$ results from contributions from $F_{vdW}$ as
well as the brush penalty. This is because our choice of $v_{hc}=v_{cc}$
leads to a $\theta $ independent $H_{eq}$. In the general case $\Delta T$
should reflect both contributions.

The analysis presented above focused on the case of short, immobile, grafted
chains that form a single-stranded helix. In this system $T_{t}$ increases
while $\Delta T$ decreases as the grafting density grows. The effect of
grafting on $\sigma$, the van der Waals attraction and the crowding induced
stretching should however play a role irrespective of the precise
specifications of the grafted layer. Additional factors may contribute when
chain mobility, multiple stranded helices and longer chains are considered.
Lateral chain mobility may give rise to in-plane phase separation driven by
the van der Waals attraction. The discussion of multiple stranded helices
should allow for the effect of loops on $\sigma$\cite{GK}. It may also be
necessary to allow for loss of translational entropy due to the formation of
a multiple helix. The ``all or nothing'' model described above is clearly
limited to brushes of short chains. A discussion of brushes formed by long
helix-forming chains should allow for helix-coil coexistence on each of the
polymers. While the discussion focused on brushes of peptides, it should be
noted that similar situation is encountered in DNA chips undergoing
hybridization\cite{DNA}. Finally, it is of interest to note that our
considerations are somewhat similar of the analysis of the coupling between
the helix-coil transition and the onset of liquid crystalline order in
peptides solutions\cite{Pincus} and in a collapsing chain\cite{Grosberg}.
From this perspective, the distinguishing features of the brush case are due
to the grafting modification of $\sigma$ and to the crowding induced chain
stretching.

\stars 
The authors benefited from instructive discussions
with M. Tirrell and D. Leckband.


\begin{thebibliography}{99}
\bibitem{Foot1}  In the vicinity of $\theta =1$ the coils no longer overlap
and some of the helices do not interact with the coils. Both $F_{cc}$ and 
$F_{hc}$ assume different forms in this regime.
\bibitem{Foot2}  The definitions of $T_{t}$ and $\Delta T$ are clearly
somewhat arbitrary. The rigorous definitions, in terms of the curvature of
the $\theta $ vs. $T$ plots, do not allow for simple analytical solutions.
\end{thebibliography}
\end{document}